\documentstyle[aps,multicol,epsf]{revtex}

\def\beginwide{
        \end{multicols} \vspace*{-0.5cm} \noindent
        \rule{3.5in}{.1mm}\rule{.1mm}{5mm} \widetext \medskip }
\def\beginwidetop{
        \end{multicols} \vspace*{-0.5cm} \noindent
        \widetext \medskip }
\def\endwide{
        \hspace*{3.35in}~\rule[-5mm]{.1mm}{5mm}\rule{3.5in}{.1mm}
        \begin{multicols}{2} \vspace*{-1.0cm} \noindent }
\def\endwidebottom{
        \begin{multicols}{2} \vspace*{-1.0cm} \noindent }

\draft

\begin{document}

\title{Self-organized branching processes: Avalanche models with dissipation}

\author{Kent B{\ae}kgaard Lauritsen,$^{1,2,}$\cite{baekgard}
	Stefano Zapperi,$^{1,}$\cite{zapperi} 
	and H. Eugene Stanley$^{1,}$\cite{hes}}

\address{$^1$Center for Polymer Studies and Department of Physics,
	Boston University, Boston, Massachusetts 02215\\
	$^2$Niels Bohr Institute, Center for Chaos and Turbulence Studies,
        Blegdamsvej 17, 2100 Copenhagen \O, Denmark}

\date{\today}

\maketitle

\begin{abstract}
We explore in the mean-field approximation
the robustness with respect to dissipation
of self-organized criticality in sandpile models. 
To this end,
we generalize a recently introduced self-organized branching process, 
and show that the model self-organizes 
not into a critical state but rather into a subcritical state:
when dissipation is present, the dynamical fixed point 
does not coincide with the critical point. 
Thus the level of dissipation acts as
a relevant parameter in the renormalization-group sense.
We study the model numerically and
compute analytically the critical exponents for the avalanche
size and lifetime distributions and the scaling exponents for
the corresponding cutoffs.

\end{abstract}


%
%
%
%
%

\begin{multicols}{2}
\bigskip

\section{Introduction}
Many driven systems in nature respond to external perturbations
by a hierarchy of avalanche events. 
This type of behavior is observed in magnetic systems 
\cite{bark}, flux lines in superconductors \cite{flux}, 
fluid flow through porous media \cite{porous}, 
microfracturing processes \cite{ae},
earthquakes \cite{gr}, and physiological phenomena \cite{lung}.
In these systems the distribution of avalanche amplitudes $s$
decays as a power law, $D(s)\sim s^{-\tau}$,
thus suggesting an analogy with
critical phenomena. Self-organized criticality 
(SOC) was proposed \cite{btw}
as a  possible framework to describe those phenomena.
Power-law scaling would emerge spontaneously due to the dynamics,
without the fine tuning of external 
parameters such as the temperature. 
Various models have been proposed with
the aim of capturing the essential features of
avalanche dynamics and self-organization.
In particular, sandpile models stimulated
an intense experimental \cite{sand,rice}, numerical \cite{nagel,grass} 
and theoretical \cite{dhar,zap,zap2} activity.

As in the case of phase transitions, mean-field theory
represents the simplest approach that gives
a qualitative description of the system.
Mean-field exponents for SOC models have been obtained in different ways 
\cite{mf,dhar2,jl,fbs,gcalda,katori,broeker-grassberger:1995},
but it turns out that their values (e.g., $\tau=3/2$)
are the same for all the models considered thus far.
This fact can easily be understood since
the spreading of an avalanche in mean-field theory 
is a branching process \cite{harris} because an avalanche can be described by 
a front of ``non-interacting particles'' that can 
either trigger subsequent activity or die out.  
The connection 
between branching processes and SOC has been investigated,
and it has been proposed that the mean-field behavior of sandpile models 
can be described by a {\em critical\/} branching process
\cite{alstr,theiler,crol,gp}. 

However, the nature of the self-organization
was not addressed by the previous approaches.
In fact the branching process is critical
only for a given value of the branching probability,
while in sandpile models there is no such tuning.
Recently, we have introduced the 
``self-organized branching process'' (SOBP) \cite{sobp}, 
a mean-field model that allows one to clarify the mechanism
of self-organization in sandpile models. Moreover,
the SOBP model can be exactly mapped onto a two-state
sandpile model in the limit $d\to\infty$, where $d$ 
is the dimension of the system.

In experiments it
can be difficult to determine whether the cutoff
in the scaling is due to finite-size effects or
due to the fact that the system is not {\em at} but rather only
{\em close to\/} the critical point. In this respect, it is
important to test the robustness of SOC behavior by
understanding which perturbations
destroy the critical properties of SOC model.

It has been shown numerically \cite{noncons1,noncons2,noncons3}
that the breaking of the conservation of 
particle numbers leads to a characteristic
 size in the avalanche distributions. 
Here we generalize the SOBP in order to allow
for dissipation and we show, in the mean-field
approximation, how the system self-organizes
in a {\em sub-critical} state.
In other words, the degree of nonconservation is a relevant parameter
in the renormalization group sense \cite{zap2}.

In section II
we derive the SOBP from a dissipative sandpile
model. In section III we study the approach to
the critical state. The critical exponents are
evaluated in section IV, and the results are
verified numerically. Section V is devoted to conclusions.

\section{Model and mean-field theory}

Sandpile models are cellular automata with an integer or 
continuous variable $z_i$ (energy) associated
with each site $i$ of a $d-$dimensional lattice.
At each time step the energy of a randomly chosen
site is increased by some amount. When the energy
on a site reaches a threshold $z_c$ the site
becomes unstable and {\em relaxes\/}
by transferring its energy to its neighbors 
according to the specific rules of the model.
In this way, a single relaxation can trigger other
relaxations, leading to the formation of an avalanche.
The boundary conditions are chosen to be open,
so avalanches that reach the boundaries
release energy outside of the system. 
After a transient, the system reaches
a steady state characterized by a balance
between the input and the output of energy.

Let us now consider a particular
sandpile model: the two-state model introduced
by Manna \cite{manna}. Energy can take only two stable
values $z_i=0$ (empty site) and $z_i=1$ (particle). 
When $z_i\geq 2$ the site relaxes, $z_i\to z_i-2$, 
and the energy of two randomly chosen neighbors 
is increased by one. This rule conserves the energy,
in this case the number of particles,
during an avalanche and leads to a stationary critical state.

Some degree of nonconservation can be introduced in the model by allowing for
energy dissipation in a relaxation event. 
In a continuous energy model this can be done
by transferring to the neighboring sites
only a fraction $(1-\epsilon)$
of the energy lost by the relaxing site \cite{noncons1}.
In a discrete energy model, such as the Manna two-state
model, one can introduce 
as the probability $\epsilon$ that the two particles
transferred by the relaxing site are annihilated \cite{noncons2}.
For $\epsilon=0$ one recovers the original two-state model.

Numerical simulations \cite{noncons1,noncons2} show
that the two ways of considering dissipation lead to the same effect:
a characteristic length is
introduced into the system and the criticality is lost.
As a result, the avalanche size distribution decays not
as a pure power law but rather as
\begin{equation}
	D(s) \sim s^{-\tau} \, h_s(s/s_c)  .
						\label{eq:D(s)-def}
\end{equation}
Here $h_s(x)$ is a cutoff function and
 the cutoff size scales as 
\begin{equation}
s_c\sim \epsilon^{-\varphi}.
\end{equation}
The size $s$ is defined as the number of sites that 
relax in an avalanche. We define the
avalanche lifetime $T$ as the number of steps comprising
an avalanche. The corresponding distribution decays as
\begin{equation}
	D(T) \sim T^{-y} \, h_T(T/T_c) ,
					\label{eq:D(T)-def}
\end{equation}
where $h_T(x)$ is another cutoff function and
$T_c$ is a cutoff that scales as 
\begin{equation}
T_c\sim \epsilon^{-\psi}.
\end{equation}
The cutoff or ``scaling'' functions $h_s(x)$ and $h_T(x)$ 
fall off exponentially
for $x\gg 1$.

To construct the mean-field theory, we
consider the model as $d\to\infty$ i.e. for an infinite dimensional lattice.
When a particle is added to an arbitrary site, 
the site will relax if a particle was
already present, which occurs with probability $p=P(z=1)$,
the probability that the site is occupied.
If a relaxation occurs, the two particles are transferred with
probability $1-\epsilon$ to two of the infinitely 
many nearest neighbors, or they are dissipated with
probability $\epsilon$.

Since $d\to\infty$ implies that the lattice coordination number  
tends to infinity, the avalanche will never visit 
the same site twice, implying that each site that receives
a particle from a neighbor relaxes with the same probability.
The avalanche process in the mean-field limit is 
a branching process. Moreover, we note that the 
branching process can be described by the {\em effective} branching
probability 
\begin{equation}
	\tilde{p} \equiv  p (1-\epsilon) , 
                                        \label{eq:p-tilde}
\end{equation}
where $\tilde{p}$ is the probability to create two new active sites.
{}From the theory of branching processes \cite{harris}, we know that there is
a critical value, $\tilde{p}=1/2$, or
\begin{equation}
	p=p_c \equiv \frac{1}{2(1-\epsilon)} , 
                                        \label{eq:p_c}
\end{equation}
such that for $p>p_c$ the
probability to have an infinite avalanche is non-zero, while for
$p<p_c$ all avalanches are finite. 
The value $p=p_c$ corresponds
to the critical case where avalanches are power law distributed.

\begin{figure}[htb]
\narrowtext
\centerline{
        \epsfxsize=8.0cm
        \epsfbox{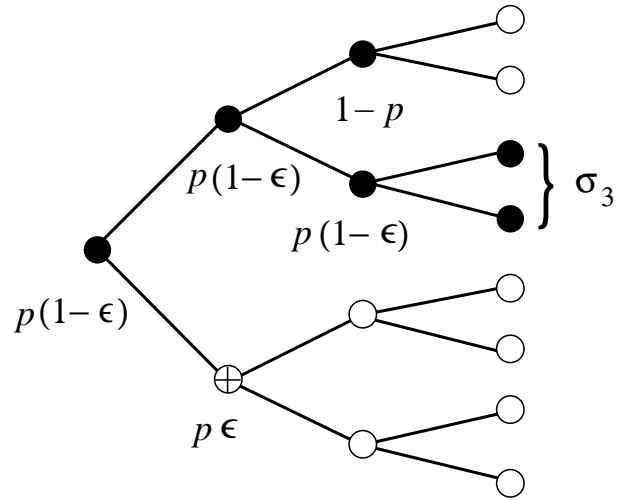}
	\vspace*{0.5cm}
}
\caption{Schematic drawing of an avalanche in a system with a
        maximum of $n=3$ avalanche generations corresponding to 
        $N=2^{n+1}-1=15$ sites. Each black site ($\bullet$) can relax
	in three different ways: 
	(i)   with probability $p(1-\epsilon)$ to two new black sites,
	(ii)  with probability $1-p$ the avalanche stops, and
	(iii) with probability $p\epsilon$ two particles are dissipated
	      at a black site, which then becomes a marked site ($\oplus$), 
	      and the avalanche stops.
        The black sites are part of an avalanche of size $s=6$,
        whereas the active sites at the boundary yield $\sigma_3(p,t)=2$.
	The total number of ``stopped'' sites are $\mu=2$, and there was
	one dissipation event such that $\kappa=2$.
        }
\label{fig:bp}
\end{figure}

Boundary conditions are important
for the process of self-organization. We can
introduce the ``boundary conditions'' 
in the mean-field theory in a natural way
by allowing for no more than $n$ generations for each avalanche. 
We can view the evolution of a single 
avalanche of size $s$ as taking place on a tree 
of $N=2^{n+1}-1$ sites (see Fig.~\ref{fig:bp}).
Note that we are not studying the model on a
Bethe lattice \cite{markosova}; 
i.e., the branching structure we are discussing
is not directly related to the geometry of the system.
The number of generations $n$ can, nevertheless, be thought of as some
measure of the linear dimension of the system.
If the avalanche reaches the boundary of the tree,
we count the number of active sites $\sigma_n$
(which in the sandpile language corresponds to the energy leaving the system),
and we expect that $p$ decreases for the next avalanche. 
If, on the other hand, the avalanche
stops before reaching the boundary, then $p$ will slightly increase.

\vspace*{0.7cm}

To make the above statements quantitative, consider
the evolution of the total number of particles $M(t)$ in the system
after each avalanche: 
\begin{equation}
	M(t+1) = M(t) + 1 - \sigma(p,t) - \kappa(p,t) .
                                                \label{eq:M(t)}
\end{equation}
Here $\sigma$ is the number of particles that leave
the system from the boundaries and $\kappa$ is the number
of particles lost by dissipation.
Since $M(t)=N P(z=1) = Np$, we obtain an
evolution equation for the parameter $p$:
\begin{equation}
	p(t+1)=p(t)+\frac{1-\sigma(p,t)-\kappa(p,t)}{N}.
                                                \label{eq:p(t)-disc}
\end{equation}
This equation reduces to the SOBP model \cite{sobp} for 
the case of no dissipation ($\kappa=0$). 
The implications of Eq.~(\ref{eq:p(t)-disc})
will be discussed in the following sections.

\section{Self-organization: the properties of the steady state}

In order to characterize the steady state of the SOBP model,
we rewrite Eq.~(\ref{eq:p(t)-disc}) in terms of
the average values of $\sigma$ and $\kappa$ indicated by angular brackets.
The average number of particles $\left<\sigma_n \right>$ leaving the
system from the boundaries in a system of $n$ generations
is computed \cite{harris} because of the recursive nature of the process:
\begin{equation}
	\left<\sigma_n(p,t)\right>=(2p(1-\epsilon))^n.
                \label{eq:sigma-split}
\end{equation}

The evaluation of the average number of particles dissipated
during an avalanche is somewhat more involved.
We can first relate the average value of $\kappa$ to the 
average number of sites $\mu$ where an avalanche does not
branch---either because of dissipation or because the site was
empty (i.e., the avalanche stops),
\begin{equation}
	\left< \kappa \right>
		= 2\left<\mu\right>\frac{p\epsilon}{p\epsilon+1-p}.
\end{equation}
The calculation of $\left< \kappa \right>$ then reduces to the
calculation of $\left<\mu\right>$.
If we denote by $\sigma_m$ the number of active sites 
at generation $m$, then $\mu$ is given by 
\begin{equation} 
	\mu=\sum_{m=0}^{n-1}\left(\sigma_m-\frac{\sigma_{m+1}}{2}\right)=
		\frac{1+s-2\sigma_n}{2},
\end{equation}
where $s=\sum_{m=0}^{n}\sigma_m$ is the total size of the avalanche.
The average value of $s$ is obtained by summing the series:
\begin{equation}
	\left< s \right> =\sum_{m=0}^{n}\left<\sigma_m\right> =
	\frac{1-(2p(1-\epsilon))^{n+1}}{1-2p(1-\epsilon)}.
\label{eq:sav}
\end{equation}

Combing Eqs.~(\ref{eq:sigma-split})-(\ref{eq:sav}),
one obtains that Eq.~(\ref{eq:p(t)-disc}) in the continuum notation becomes 

\beginwide
\begin{equation}
	\frac{d p}{d t}=\frac{1}{N}
	\left( 1-(2p(1-\epsilon))^n-\frac{p\epsilon}{p\epsilon+1-p}
		\left[1+\frac{1-(2p(1-\epsilon))^{n+1}}{1-2p(1-\epsilon)}
			- 2(2p(1-\epsilon))^n
		\right]
	\right)
	+\frac{\eta(p,t)}{N}.
					\label{eq:p(t)-cont}
\end{equation}

\endwide
In Eq.~(\ref{eq:p(t)-cont}), we introduced the function $\eta(p,t)$ to 
describe the fluctuations around the average values of $\sigma$ and $\kappa$.
We have shown numerically that the effect of
this ``noise'' term  is vanishingly small in the limit
$N\to\infty$. 

\begin{figure}[htb]
\narrowtext
\centerline{
        \epsfxsize=7.0cm
        \epsfbox{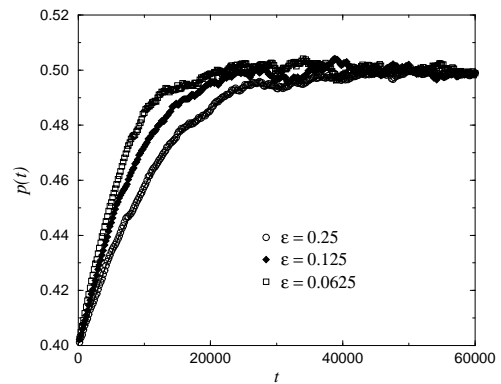}
        }
\caption{The value of the control parameter $p(t)$
	as a function of time for a system with 
        different levels of dissipation.
        After a transient, $p(t)$ reaches its 
        fixed-point value $p^*=1/2$ and fluctuates around it with 
        short-range time correlations.
        }
\label{fig:p}
\end{figure}

Without the noise term 
we can study the fixed points of Eq.~(\ref{eq:p(t)-cont}). 
We find that there is only one fixed point, 
\begin{equation}
        p^{*}=1/2   ,
                        \label{eq:p*}
\end{equation}
independent of the value of $\epsilon$; the
corrections to this value are of the order $O(1/N)$.
By linearizing Eq.~(\ref{eq:p(t)-cont}), we find that the fixed point is
attractive.
This result implies that the SOBP model self-organizes
into a state with $p=p^*$.  
In Fig.~\ref{fig:p} we show the value of $p$ as a function of time for
different values of the dissipation $\epsilon$.
We find that independent of the initial conditions
after a transient $p(t)$ reaches the self-organized steady-state
described by the fixed point value
$p^*=1/2$ and fluctuates around it with short-range correlations
(of the order of one time unit).
The fluctuations around the critical value decrease with the system size
as $1/N$. Thus it follows that in the limit $N \to \infty$
the distribution $\phi(p)$ of $p$ approaches a delta function 
\begin{equation}
	\phi(p) \sim \delta(p - p^*).  
					\label{eq:phi(p)}
\end{equation}

\begin{figure}[htb]
\narrowtext
\centerline{
        \epsfxsize=7.0cm
        \epsfbox{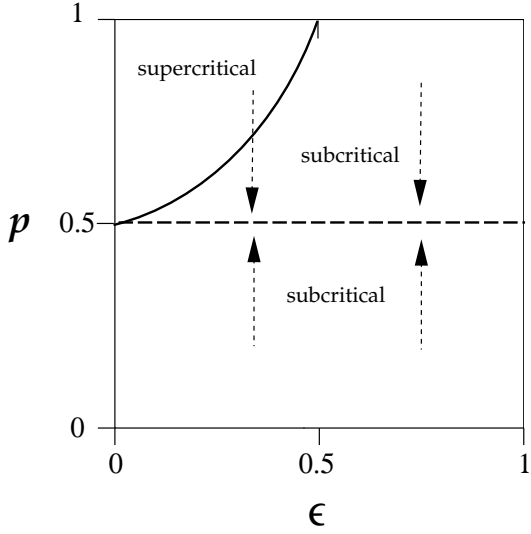}
	\vspace*{0.5cm}
        }
\caption{Phase diagram for the SOBP model with dissipation.
	The dashed line shows the fixed points $p^*=1/2$ of the
	dynamics, with the flow being indicated by the arrows. 
	The solid line shows the critical points, 
	cf.\ Eq.~(\protect\ref{eq:p_c}).
        }
\label{fig:flow}
\end{figure}

By comparing the fixed point value (\ref{eq:p*}) with the critical value
(\ref{eq:p_c}), we obtain that in the presence of
dissipation ($\epsilon >0$) the self-organized steady-state of the system
is {\em subcritical}.
Fig.~\ref{fig:flow}
is a schematic picture of the phase space of the model, including
the line $p=p_c$ of critical behavior (\ref{eq:p_c}) and the 
line $p= p^*$ of fixed points (\ref{eq:p*}). 
These two lines intersect only for $\epsilon=0$.

\vspace*{0.5cm}

\section{Critical exponents}
\label{sec:crit-exps}

In this section, we study the critical properties of the model.
In the limit $n \gg 1$ we obtain analytical results
for the avalanche and lifetime distributions for any value of 
$\tilde{p}$, the effective branching probability defined in
Eq.~(\ref{eq:p-tilde}).
We show that the critical branching process 
with $\tilde{p}=1/2$ (obtained when $\epsilon=0$)
correctly reduces to the mean-field exponents $\tau=3/2$ and $y=2$.

\subsection{Generating functions}

The quantities $P_n(s,\tilde{p})$ and $Q_n(\sigma,\tilde{p})$
are defined to be the probabilities of an avalanche of size
$s$ and boundary size $\sigma$ respectively,
in a system with $n$ generations.
The corresponding generating functions are defined by \cite{harris}
\begin{mathletters}
\label{eq:hg_def}
\begin{equation}
        f_n(x,\tilde{p}) \equiv \sum_{s} P_n(s,\tilde{p}) x^s   ,
						\label{eq:f_n-def}
\end{equation}
\begin{equation}
	g_n(x,\tilde{p}) \equiv \sum_{\sigma} Q_n(\sigma,\tilde{p}) x^\sigma .
						\label{eq:g_n-def}
\end{equation}
\end{mathletters}
{}From the hierarchical structure of the branching process, it is
possible to write down recursion relations for $P_n(s,\tilde{p})$ and
$Q_n(\sigma,\tilde{p})$, from which we obtain \cite{harris}
\begin{mathletters}
\label{eq:fg_n+1}
\begin{equation}
        f_{n+1}(x,\tilde{p}) 
	= 
	x \left[ (1-\tilde{p}) + \tilde{p} f^{2}_{n}(x,\tilde{p}) \right]  
                                                \label{eq:f_n+1}
\end{equation}
and
\begin{equation}
        g_{n+1}(x,\tilde{p}) 
	=
	(1-\tilde{p}) + \tilde{p} g^{2}_{n}(x,\tilde{p})  ,
                                                \label{eq:g_n+1}
\end{equation}
\end{mathletters}
where $f_0(x,\tilde{p})=g_0(x,\tilde{p})=x$.

\subsection{Avalanche size distribution}

The solution of Eq.~(\ref{eq:f_n+1}) in the limit $n \gg 1$ is given by
\begin{equation}
	f(x,\tilde{p}) 
	= 
	\frac{1-\sqrt{1-4x^2\tilde{p}(1-\tilde{p})} \, }{2x\tilde{p}} .
                                                \label{eq:f*}
\end{equation}
We expand Eq.~(\ref{eq:f*}) as a series in $x$, and by comparing with
the definition~(\ref{eq:f_n-def}), we obtain for sizes such that
$1 \ll s \lesssim n$ \cite{small-s}
\begin{equation}
        P_n(s,\tilde{p}) = 
		\frac{\sqrt{2(1-\tilde{p})/\pi \tilde{p}}}{s^{3/2}} \, 
	 	\exp\left( -s / s_c(\tilde{p}) \right)  .
                                                         \label{eq:P_n}
\end{equation}
The cutoff $s_c(\tilde{p})$ is given by
\begin{equation}
	s_c(\tilde{p}) = -\frac{2}{\ln 4\tilde{p}(1-\tilde{p})} .
                                                     \label{eq:cutoff}
\end{equation}
For avalanches with $n \lesssim s \lesssim N$ it is possible to
use a Tauberian theorem \cite{feller,asmussen-hering,weiss}, and show
that $P_n(s,\tilde{p})$ will decay exponentially.

The next step is to calculate the avalanche distribution
$D(s)$ for the SOBP model. This can be calculated 
as the average value of $P_n(s,\tilde{p})$ with respect to the probability
density $\phi(p)$, i.e., according to 
\begin{equation}
        D(s) = \int_{0}^{1} dp \, 
			\phi(p) \, P_n(s,\tilde{p})   .
                                                \label{eq:D(s)-int}
\end{equation}
Since the simulation results show that $\phi(p)$ for $N \gg 1$
approaches the delta function $\delta(p-p^*)$
[cf.\ Eq.~(\ref{eq:phi(p)})], expression (\ref{eq:D(s)-int}) reduces to
\begin{equation}
        D(s) = \left. 
		P_n(s,\tilde{p})
		\right|_{\tilde{p}=p^*(1-\epsilon)} .
                                                \label{eq:D(s)-2}
\end{equation}
As a result we obtain the distribution
\begin{equation}
        D(s) = \sqrt{\frac{2}{\pi}\,} \,
                \frac{1+\epsilon+ \ldots}{s^{\tau}}
                \exp\left( - s/s_c(\epsilon) \right) .
                                                \label{eq:D(s)}
\end{equation}
We can expand $s_c(\tilde{p})$ in $\epsilon$ with the result
\begin{equation}
        s_c(\epsilon) \sim \frac{2}{\epsilon^\varphi}  , 
	~~~~~~ \varphi = 2 .
				\label{eq:phi=2}
\end{equation}
Furthermore, the mean-field exponent
for the critical branching process is obtained setting $\epsilon=0$, i.e.,  
\begin{equation}
	\tau=3/2 .
				\label{eq:tau}
\end{equation}

\begin{figure}[htb]
\narrowtext
\centerline{
        \epsfxsize=7.0cm
        \epsfbox{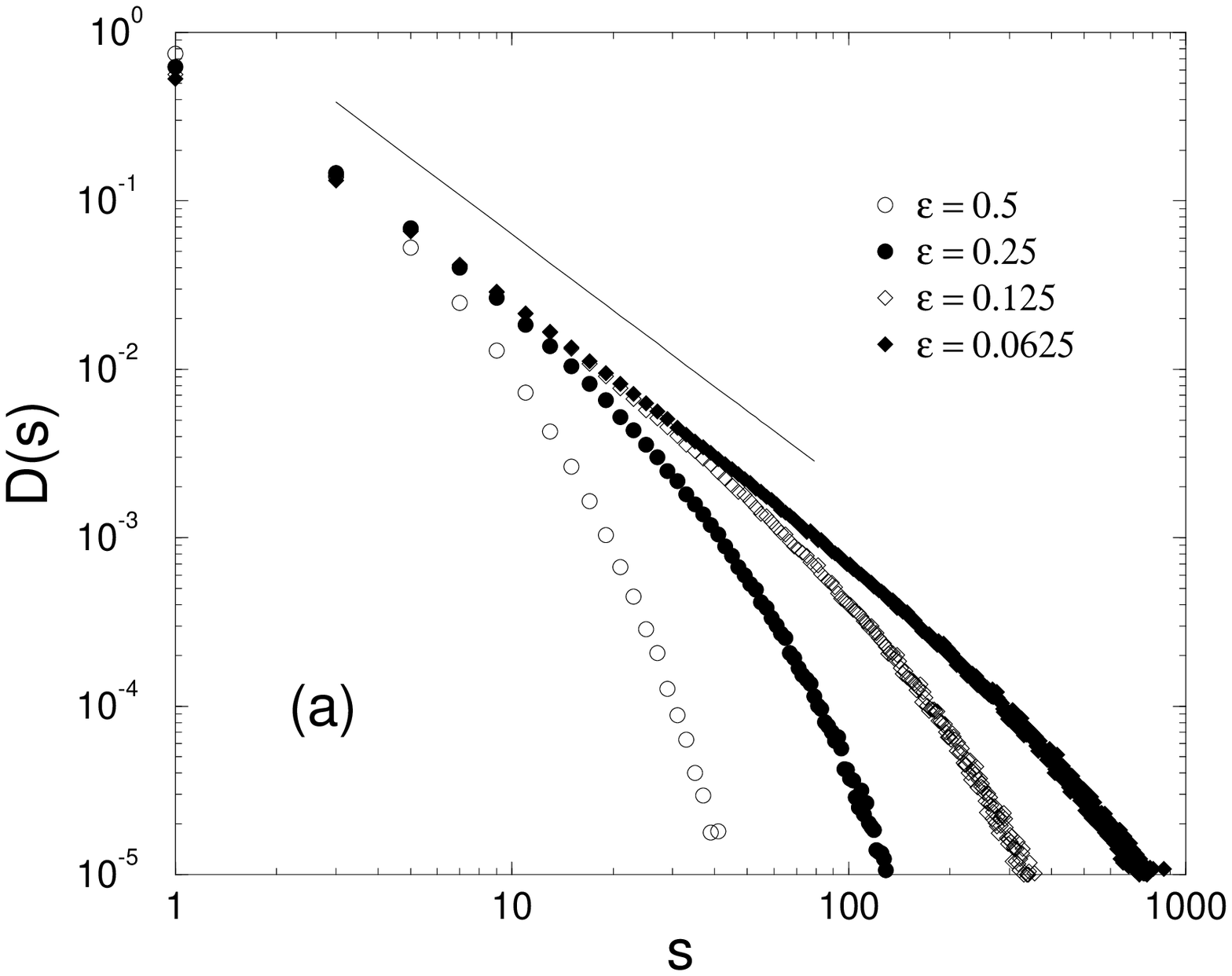}
        }
\centerline{
        \epsfxsize=7.0cm
        \epsfbox{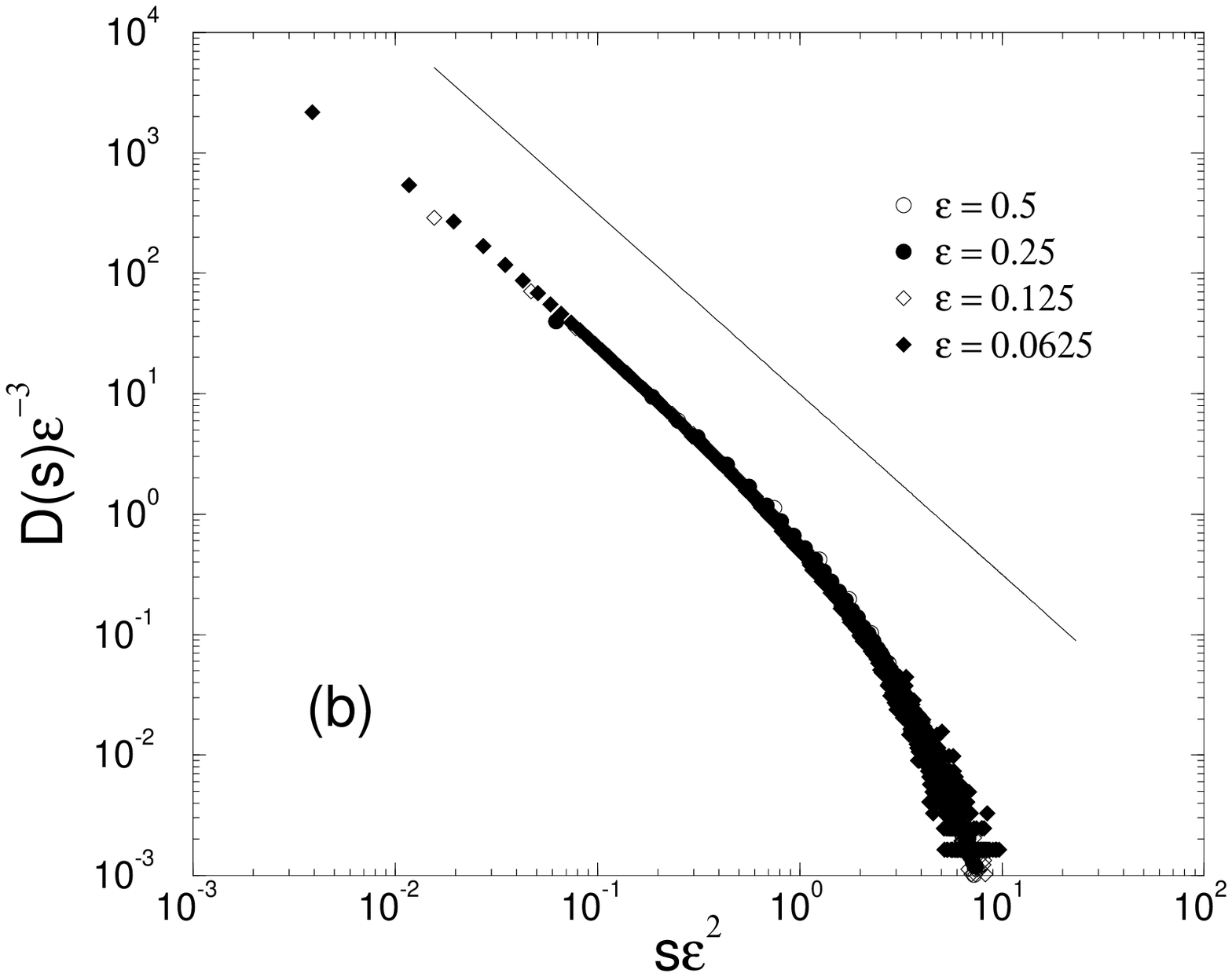}
        }
\caption{(a) Log-log plot of the avalanche distribution $D(s)$ 
        for different levels of dissipation. 
        A line with slope $\tau=3/2$ is plotted for reference,
        and it describes the behavior of the data for intermediate $s$ values,
        cf.\ Eq.~(\protect\ref{eq:D(s)}).
        For large $s$, the distributions fall off exponentially.
	(b) Data collapse produced by Eq.~(\protect\ref{eq:D(s)}).
        }
\label{fig:ds}
\end{figure}

These results are in excellent agreement with the simulation of $D(s)$ for
the SOBP model (cf.\ Fig.~\ref{fig:ds}).
The deviations from the power-law behavior~(\ref{eq:D(s)})
are due to the fact that Eq.~(\ref{eq:P_n}) is only valid for
$1 \ll s \lesssim n$ \cite{small-s}.

\subsection{Lifetime distribution}

The avalanche lifetime distribution $D(T)$ is defined,
for the model, as
the probability to obtain an avalanche which spans $m$ generations;
here, we identify $m$ with the time $T$.
It follows that
\begin{equation}
	P(m,\tilde{p}) = Q_{m+1}(\sigma=0,\tilde{p}) 
				- Q_{m}(\sigma=0,\tilde{p})  .
							\label{eq:D(T)-Pm}
\end{equation}
As for the avalanche distribution $D(s)$ we have that
$D(T)=P(m=T,\tilde{p})$ evaluated for $\tilde{p}=p^*(1-\epsilon)$.

For $\tilde{p}=1/2$ we use the general result \cite{harris}
\begin{equation}
        \frac{1}{1-Q_m(\sigma =0, \tilde{p})} 
		= 1 + m\tilde{p} + O(\ln m), 
	~~~~m \gg 1,
                                                        \label{eq:f_m(0)}
\end{equation}
and obtain
\begin{equation}
	P(T, \tilde{p}) = \frac{\tilde{p}^{-1}}{T^2} 
		\left( 1 + O(\ln T / T) + \ldots \right)  .
							\label{eq:D(T)-2}
\end{equation}
Note the strong correction to scaling to $D(T)$ in this case.
For $\tilde{p}<1/2$ we find \cite{harris}
\begin{equation}
	1 - Q_m(\sigma = 0,\tilde{p}) \sim c_1 (2\tilde{p})^m  ,
\end{equation}

\begin{figure}[htb]
\narrowtext
\centerline{
        \epsfxsize=7.0cm
        \epsfbox{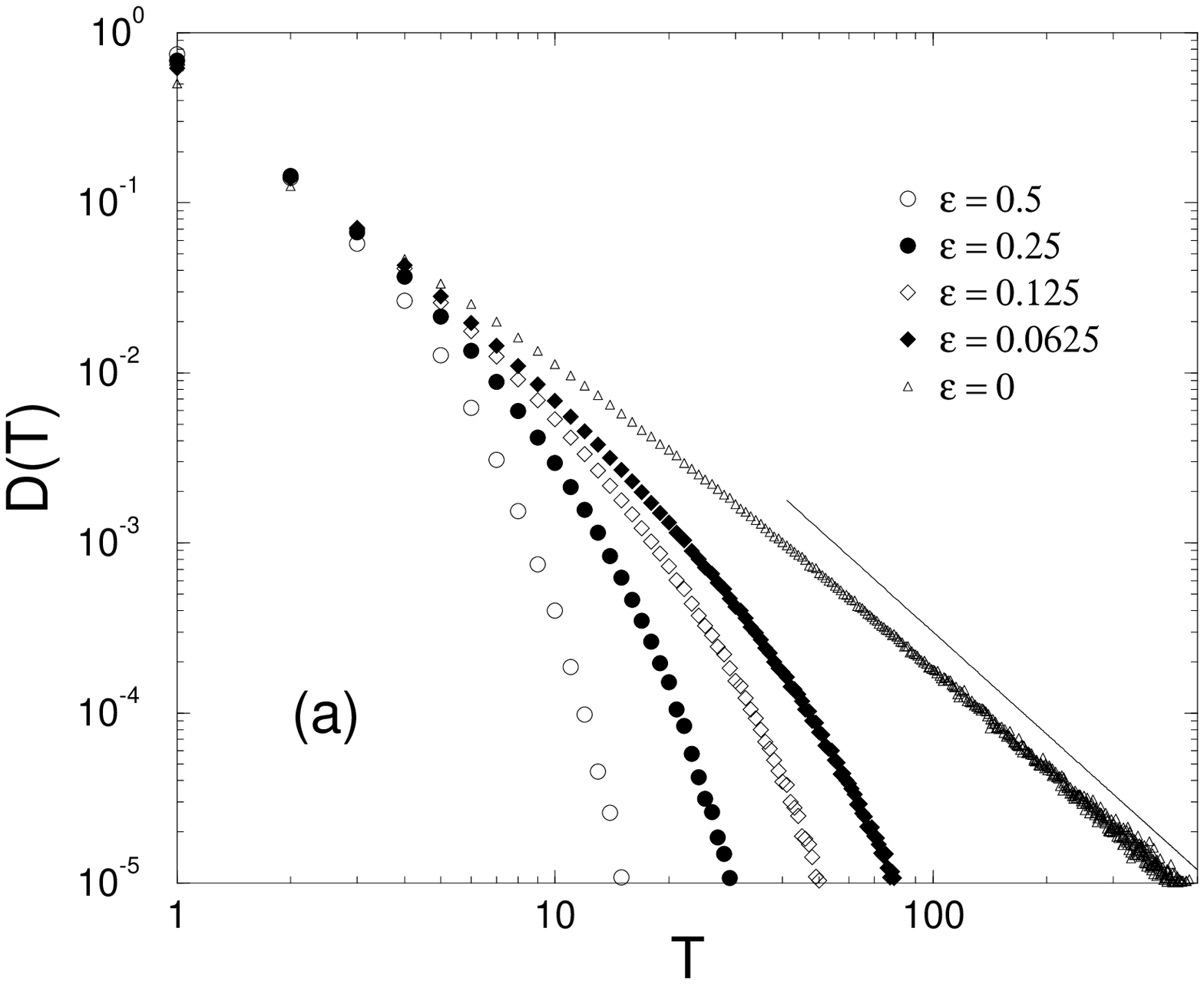}
        }
\centerline{
        \epsfxsize=7.0cm
        \epsfbox{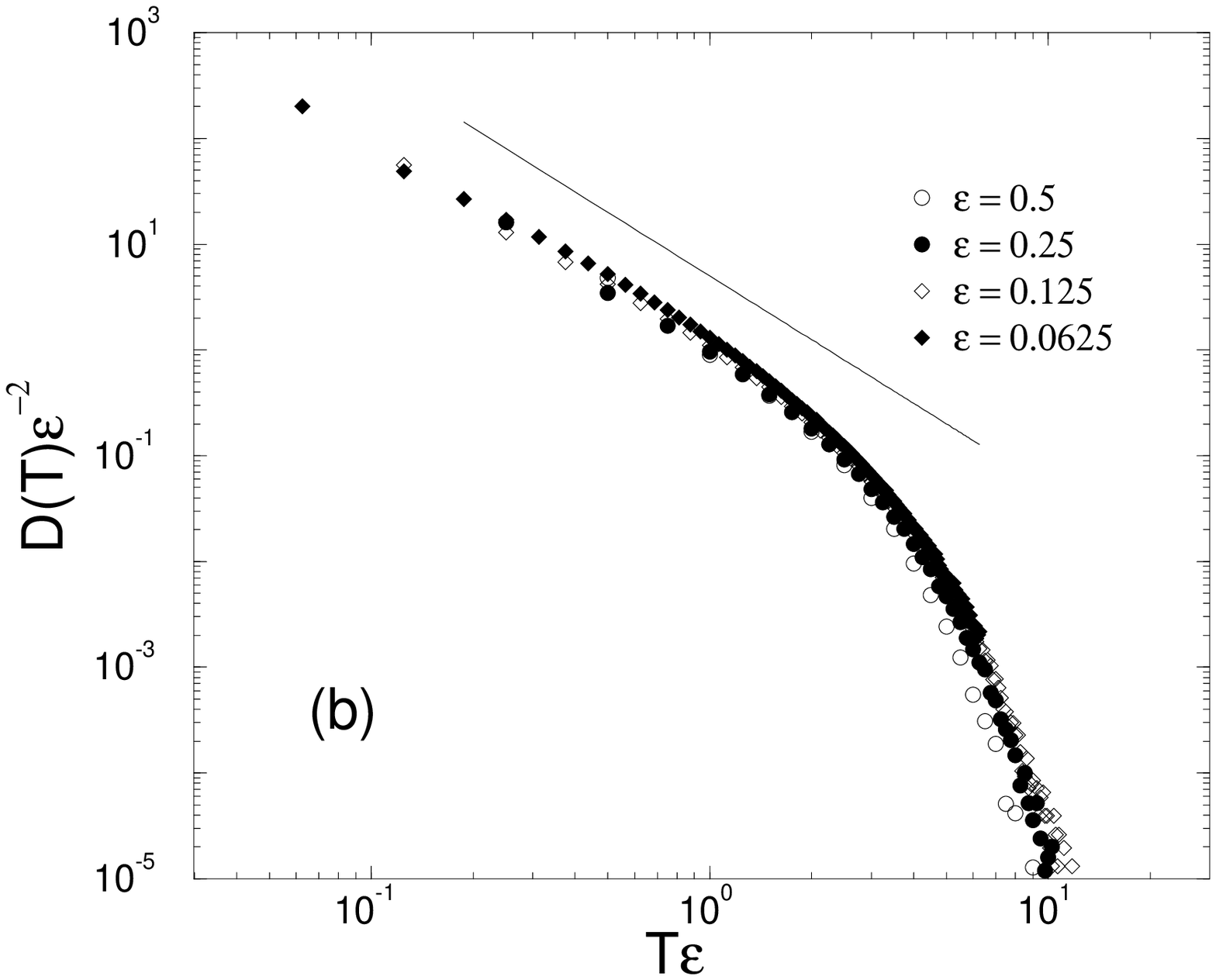}
        }
\caption{(a) Log-log plot of the lifetime distribution $D(T)$ 
        for different levels of dissipation. 
	A line with slope $y=2$ is plotted for reference. 
	Note the initial deviations from the power law for $\epsilon=0$
	due to the strong corrections to scaling,
	cf.\ Eq.~(\protect\ref{eq:D(T)-2}).
	(b) Data collapse produced by Eq.~(\protect\ref{eq:D(T)}).
        }
\label{fig:dt}
\end{figure}

\vspace*{0.5cm}

\noindent
for $m \gg 1$, where $c_1 > 0$ is an unknown constant.
This expression yields $D(T) \sim \epsilon \exp(-n\epsilon)$.

We can combine the above results in the scaling form
\begin{equation}
        D(T) \sim T^{-y} \exp(-T/T_c)  , 
                                                   \label{eq:D(T)}
\end{equation}
where
\begin{equation}
	T_c \sim \epsilon^{-\psi},  ~~~~~~ \psi = 1 .
					\label{eq:psi=1}
\end{equation}
The lifetime exponent $y$ was defined in Eq.~(\ref{eq:D(T)-def}), 
wherefrom we obtain the mean-field result
\begin{equation}
        y = 2. 
                                                \label{eq:y=2}
\end{equation}
In Fig.~\ref{fig:dt}, we show lifetime distributions
for different values of $\epsilon$,
together with the data collapse produced by Eq.~(\ref{eq:D(T)}).

\section{Discussion and conclusions}

We have studied the effect of dissipation in the dynamics
of the sandpile model in the mean-field limit ($d\to\infty$).
In this limit, the dynamics of an avalanche is described 
by a branching process. We have derived an evolution equation for
the branching probability that generalizes the
self-organized branching process (SOBP) introduced in Ref.\ \cite{sobp}.
By analyzing this evolution equation, we have shown that there is
a single attractive fixed point which in the
presence of dissipation is not a critical point. The 
level of dissipation $\epsilon$ therefore acts as a relevant
parameter for the SOBP model. We have determined analytically the critical
exponents describing the scaling of the characteristic size
with $\epsilon$ and the form of the avalanche distributions, and
numerically verified the above results.

These results prove, in the mean-field limit, that 
criticality in the sandpile model is lost when dissipation
is present. It would be interesting to use a similar approach 
for other forms of perturbations. In particular
it has been shown for other SOC models that the presence
of a non-zero temperature \cite{temp} or of a non-zero
driving rate \cite{ff} are relevant perturbations
leading to a non-critical steady state.

Finally, we discuss the relations between the SOBP
model and the {\em simplest possible SOC system\/} recently
introduced by Flyvbjerg \cite{flyvbjerg:1996}. 
The minimal definition of SOC, as a medium in which externally
driven disturbances propagate leading to a stationary
critical state, is well exemplified by the
SOBP model. The disturbance is described by the
branching process and the medium by the evolution equation
for the density of particles in the system [Eq.~(\ref{eq:p(t)-disc})].
The example given by Flyvbjerg, being a two-state 
random-neighbor sandpile model, differs 
from the SOBP \cite{sobp}
in the way open boundary conditions are imposed.

\section*{Acknowledgments}

K.~B.~L.\ acknowledges the support from the Danish Natural Science
Research Council. The Center for Polymer Studies is supported by NSF.

\end{multicols}


\begin{thebibliography}{99}
\bibitem{small-s}
        For small $s$ we have the exact results
        $P_n(1,p)=1-p$, $P_n(3,p)=p(1-p)^2$, $P_n(5,p)=2p^2(1-p)^3$,
        and so forth.
\end{thebibliography}
\end{document}